# The Big Bang Was Not That Bright

Yaron Sheffer
Postdoc Emeritus in Astronomy
yaron42sheffer@gmail.com

**Abstract**
A recent arXiv manuscript claims that a cosmic background radiation with a black body temperature of $T_{BB}$ ~ 500 K (440 F) was just barely visible to human eyes, thus fixing the onset of the Dark Ages at about 5 million years post recombination. This claim presents an insurmountable biophysical challenge, since even hotter bodies, such as 450 F pizzas, do not seem to be glowing in the dark. As volunteer referees we show that this claim is the result of employing an incorrect assumption. Via a corrected analysis we find that the Dark Ages must have had a significantly earlier start. A second, more descriptive claim, that a cosmic background radiation with $T_{BB}$ of 1545 K was as blinding to humans as is our own Sun, is based on the same assumption and may have to be revised.

**Introduction**
In the concordance model, the cosmic background radiation (CBR) decoupled from matter 380 ky after the Big Bang, and has been expanding and cooling down ever since. Before the formation of the first stars, the Universe was going through "Dark Ages" (DAs) until an age of ~150 My. The authors [1] analyze human visual perception of the young CBR and conclude that the DAs did not start until ~5 My post recombination (PR). It will be shown that the DAs must have started earlier than the authors' finding, following their dramatic underestimation of the threshold level of darkness perception.

**Making Analysis Better**
The authors take their own candle-based darkness limit for the human eye and then assume that the same power level of $1.8 \times 10^{-14}$ W can be applied to the isotropic CBR when viewed over the entire retinal field of view (FOV). Alas, this assumption contradicts basic visual biophysics: the eye detects light using individual detectors (rods and cones), not via summation over the entire surface of the retina. Spreading the power from a point source over the FOV guarantees that the CBR will be exceedingly undetectable by any of the rods and cones of the human eye. Undetectable by how much? The candle light, having a solid angle of $7.3 \times 10^{-12}$ sr, would become diluted by a factor of about $10^{12}$(!) over the 5.2 sr of the retina. Thus the threshold of darkness is underestimated by some 30 stellar magnitudes, presenting a detection impossibility even for the Hubble Space Telescope, and assuring that the CBR would be completely invisible to the human eye already before 5 My PR.

The methodology employed by the authors involves matching the diluted power with a CBR of proportionately reduced brightness at a lower value of $T_{BB}$, in this case ~500 K [2]. For the correct analysis one should keep the original undiluted power level without varying any of the parameters such as the solid angle of the source. This can be straightforwardly accomplished with a CBR having the original source $T_{BB}$ (assumed 1400 K for the candle) and providing a full retinal glow-in-the-dark power level of ~$10^{-2}$ W. Based on the authors' Fig. 1 and equations, such a warmer CBR occurs as early as ~1 My PR, rather than 5 My, at a redshift (1+z) ~ 510.

But before adopting this new and improved solution for the start of the DAs, we must attend to one additional complication.

**Making Analysis Better Again**
The power from unresolved sources, such as a far away candle or a star, falls on the retina in a circular area (Airy disk) that defines the eye's resolution element. Since this is the minimum solid angle that the eye can resolve, all power detection calculations should take this minimum into account. For a low-light-level pupil 8 mm across, this area is 17.5" in radius, with a solid angle of $2.3 \times 10^{-8}$ sr. Spreading this Airy disk over the entire retinal area of 5.2 sr would result in a dilution factor of $2.3 \times 10^8$, which is not as dramatic as $\sim 10^{12}$. This improved factor means that the incorrect methodology would underestimate the CBR darkness threshold by "only" 21 stellar magnitudes, still a challenging feat for the Hubble Space Telescope.

Again, we shall use the authors' Fig. 1 and their equations to estimate an earlier ending time for the Cosmic Dusk. The corrected full retinal power level for the onset of darkness becomes $\sim 3 \times 10^{-6}$ W. It corresponds to $T_{BB} \sim 900$ K [3]. The Universe cooled down to this value at a redshift of $(1+z) \sim 330$, when it was $\sim 2.5$ My old, indicating that the DAs essentially got going around 2 My PR, still significantly earlier than the 5 My value found by the authors.

**From Darkness to Blindness**
Employing an identical methodology, the authors take the visually blinding power of the solar disk and then spread it over the full retina to find its CBR equivalent at 1545 K. This time, the dilution ratio is 5.2 sr over the solar solid angle, or 77,000. When thus diluted, each sun-sized solid angle is no brighter than 7 full moons, clearly not posing any local blinding qualities. But should a whole retina covered by 77,000 extra-bright full moons be considered a global blindness limit? We do not know the answer to this question, but if not, then neither should a CBR at 1545 K. For the sake of consistency with the analysis of darkness, we should define the CBR as truly blinding if it reproduces the brightness of the solar disk, i.e., when it is as hot as the solar $T_{BB}$ of 5780 K. The only time that the CBR was that hot happened before the time of recombination, leaving the entire timeline of PR cooling with brightness levels that are well below that of the solar photosphere.

**Concluding Remarks**
Granted, the Dark Ages could not have started exactly at recombination, when light was still plentiful (yet perhaps not as blinding as imagined by the authors). But we can now "see" that the Universe was already utterly black, as far as human vision is concerned, well before the time of 5 million years post recombination. Our contribution illustrates how incorrect assumptions can significantly impact the final results. Other editorial and numerical issues identified by us in the same manuscript can be found in a more exhaustive version of this volunteer-referee report.

**Notes**
[1] "How Bright Was the Big Bang?", arXiv:1801.03278
[2] A $10^{12}$ factor in brightness can be accommodated by a factor of merely 3 in $T_{BB}$ thanks to the exponential behavior of the BB function.
[3] Faintest glow in the dark happens at $\sim 700$ K, cf. Wikipedia: Red heat.